\RequirePackage{fix-cm}
\documentclass[twocolumn]{svjour3}          % twocolumn
\smartqed  % flush right qed marks, e.g. at end of proof
\usepackage[utf8]{inputenc}  %arXivAdmin
\usepackage{graphicx,import}
\usepackage{amsmath}
\usepackage{amssymb}
\usepackage{floatrow} 
\usepackage{subfig}
\usepackage{lipsum}
\usepackage{anyfontsize}
\usepackage{mathtools}
\usepackage{cuted}
\usepackage[export]{adjustbox}
\usepackage{bibentry}
\usepackage{ulem}
\usepackage[breaklinks=true,colorlinks=true,linkcolor=blue,urlcolor=blue,citecolor=blue]{hyperref}
\newcommand{\doubleunderline}[1]{\underline{\underline{#1}}}
\newcommand{\norm}[1]{\left\lVert#1\right\rVert}

\usepackage{float}
\floatstyle{plaintop}
\restylefloat{table}
\journalname{Cognitive Computation}
\begin{document}

\title{Limitations of the recall capabilities in delay based reservoir computing systems%\thanks{Grants or other notes
}


\author{Felix Köster\and Dominik Ehlert\and Kathy Lüdge}

%\authorrunning{Felix Köster} % if too long for running head

\institute{Felix Köster (corresponding author) \at
            Institut f\"ur Theoretische Physik, Technische Universit\"at Berlin, Berlin 10623 Germany
              Straße des 17. Juni 135 \\
              Tel.: +49-30-314-24254\\
              \email{f.koester@tu-berlin.de}           %  \\
            \and
           Dominik Ehlert \at
           Institut f\"ur Theoretische Physik, Technische Universit\"at Berlin, Berlin 10623 Germany
              Straße des 17. Juni 135 \\
              \email{dominik.ehlert@posteo.de}           %  \\
           \and
           Kathy Lüdge \at
           Institut f\"ur Theoretische Physik, Technische Universit\"at Berlin, Berlin 10623 Germany
              Straße des 17. Juni 135 \\
              Tel.: +49-30-314-23002\\
              \email{kathy.luedge@tu-berlin.de}           %  \\
}

\date{Received: date / Accepted: date}

\maketitle

\begin{abstract}
\textbf{Objectives:} 
We analyze the memory capacity of a delay based reservoir computer with a Hopf normal form as nonlinearity and numerically compute the linear as well as the higher order recall capabilities. A possible physical realisation could be a laser with external cavity, for which the information is fed via electrical injection.
\textbf{Methods:}
A task independent quantification of the computational capability of the reservoir system is done via a complete orthonormal set of basis functions.
\textbf{Results:}
Our results suggest that even for constant readout dimension the total memory capacity is dependent on the ratio between the information input period, also called the clock cycle, and the time delay in the system.
\textbf{Conclusions:} 
Optimal performance is found for a time delay about 1.6 times the clock cycle
%We analyze the memory capacity of a delay based reservoir computer with a Hopf normal form as nonlinearity and numerically compute the linear as well as the higher order recall capabilities. A possible physical realisation could be a laser with external cavity or other oscillating systems with delay. A task independent quantification of the computational capability of the reservoir system is done via a complete orthonormal set of basis functions. Our results suggest that even for constant readout dimension the total memory capacity is dependent on the ratio between the information input period, also called the clock cycle, and the time delay in the system. A result of ours indicates that for optimal performance the time delay should be about 1.6 times the clock cycle.
\end{abstract}

\section{Introduction}
\label{Introduction}

Reservoir computing is a machine learning paradigm \cite{JAE01} inspired by the human brain \cite{MAA02}, which utilizes the natural computational capabilities of dynamical systems. 
As a subset of recurrent neural networks it was developed to predict time-dependent tasks with the advantage of a very fast training procedure. 
Generally the training of recurrent neural networks is connected with high computational cost resulting e.g. from connections that are correlated in time. 
Therefore, problems like the vanishing gradient in time arise \cite{HOC98}. 
Reservoir computing avoids this problem by training just a linear output layer, leaving the rest of the system (the reservoir) as it is. 
Thus, the inherent computing capabilites can be exploited. One can divide a reservoir into three distinct subsystems, the input layer, which corresponds to the projection of the input information into the system, the dynamical system itself that processes the information, and the output layer, which is a linear combination of the system's states trained to predict an often time-dependent task. \\
Many different realisations have been presented in the last years, ranging from a bucket of water \cite{FER03} over field programmable gate arrays (FPGAs) \cite{ANT16} to dissociated neural cell cultures \cite{DOC09}, being used for satellite communications \cite{BAU15}, real-time audio processing \cite{KEU17,SCA16}, bit-error correction for optical data transmission \cite{ARG17}, amplitude of chaotic laser pulse prediction \cite{AMI19} and cross-predicting the dynamics of an injected laser \cite{CUN19}.
Especially opto-electronic \cite{LAR12,PAQ12} and optical setups \cite{BRU13a,VIN15,NGU17,ROE18a,ROE20} were frequently studied because their high speed and low energy consumption makes them preferable for hardware realisations. \\
The interest in reservoir computing was refreshed when Appeltant et al. showed a realisation with a single dynamical node under influence of feedback \cite{APP11}, which introduced a time-multiplexed reservoir rather than a spatially extended system.
A schematic sketch is shown in Fig. \ref{fig:sketch}.
In general the delay architecture slows down the information processing speed but reduces complexity of the hardware.
Many neuron based, electromechanical, opto-electronic and photonic realisations \cite{ORT17a,DIO18,BRU18a,CHE19c,HOU18,SUG20} showed the capabilites from time-series predictions \cite{BUE17,KUR18} over an equalization task on nonlinearly distorted signals \cite{ARG20} up to fast word recognition \cite{LAR17}. More general analysis showed the general and task-independent computational capabilities of semiconductor lasers \cite{HAR19}. A broad overview is given in \cite{BRU19,SAN17a}. \\
In this paper we perform a numerical analysis of the recall capabilities and the computing performance of a simple nonlinear oscillator, modelled by a Hopf normal form, with delayed feedback. We calculate the total memory capacity as well as the linear and nonlinear contributions using the method derived by Dambre et al. in \cite{DAM12}. \\
The paper is structured as follows.
First, we shortly explain  the concept of time-multiplexed reservoir computing and give a short overview of the method used for calculating the memory capacity.
After that we present our results and discuss the impact of the delay time on the performance and the different nonlinear recall contributions.
\section{Methods}
\label{Methods}
\begin{figure}
	\centering
	\def\svgwidth{\textwidth}
	\import{./}{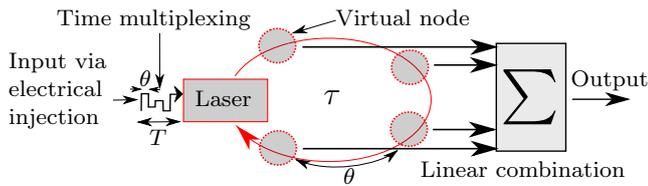}
	\caption[schematic]{Schematic sketch of time-multiplexed reservoir computing scheme. The input is preprocessed by multiplication with a mask that induces the time-multiplexing and is then electrically injected. The laser in our case is governed by a Hopf normal form. The output dimension of the system is in this example 4.}
	\label{fig:sketch}
\end{figure}
Traditionally, reservoir computing was realised by randomly connecting nodes with simple dynamics (for example the $\tanh$-function \cite{JAE01}) to a network, which was then used to process information. The linear estimator of the readouts is then trained to approximate a target, e.g. predict a time-dependent task. The network thus transforms the input into a high dimensional space in which the linear combination can be used to seperate different inputs, i.e. to classify the given data. \\
In the traditional reservoir computing setup a reaction from the system $\vec{s}_n = (s_{1n},s_{2n},\ldots,s_{Mn}) \in \mathbb{R}^M$ is recorded together with the corresponding input $u_n$ and the target $o_n$. In this case $n$ is the index for the $n$-th input-output training datapoint, ranging from $1$ to $N$, and $M$ is the dimension of the measured system states.
The goal for the reservoir computing paradigm is to approximate the target $o_n$ as close as possible with linear combinations of the states $\vec{s}_n$ for all input-output pairs $n$, meaning that $\sum_{m=1}^{M}w_{m} s_{mn} = \hat{o}_{n} \approx o_{n}$ for all $n$, where $\vec{w} = (w_{1}, w_{2},\ldots,w_{M}) \in \mathbb{R}^{M}$ are the weights to be trained.
We want to find the best solution for
\begin{align}
    \doubleunderline{s} \cdot \vec{w} \approx \vec{o},
\end{align}
where $\doubleunderline{s} \in \mathbb{R}^N \times \mathbb{R}^M$ is the state matrix defined by all system state reactions $\vec{s}_n$ to their corresponding inputs $u_n$, $\vec{w}$ are the weights to train and $\vec{o} \in \mathbb{R}^N$ is the vector of targets to be approximated.
This is equivalent to a least square problem which is analytically solved by \cite{WIL16}
\begin{align}
    \vec{w} = (\doubleunderline{s}^T \doubleunderline{s})^{-1} \doubleunderline{s}^T \vec{o}.
\end{align}
The capability of the system to approximate the target task can be quantified by the normalized root mean square difference between the approximated answers $\hat{o}_n$ and the targets $o_n$
\begin{align}
    \text{NRMSE} = \sqrt{\frac{\sum\limits_{n=1}^{N}(o_{n} - \hat{o}_{n})^2}{N \cdot var(\vec{o})}} ,
\end{align}
where NRMSE is the normalized root mean-square error of the target task with $var(\vec{o})$ being the variance of the target values $\vec{o}=(o_1,o_2,\ldots,o_N)$ and N the number of sample points.
An NRMSE of 1 indicates that the system is not capable of approximating the task better than approximating the mean value, a value NRMSE=0 indicates that it is able to compute the task perfectly.
For a successful operation $N \gg M$ needs to be fulfilled, where $M$ is the number of output weights $w_{m}$ and $N$ is the number of training data points. This corresponds to a training data set of size $N$ being significantly bigger than the possible output dimension $M$ to prevent overfitting.\\
Appeltant et$.$ al$.$ introduced in \cite{APP11} a time multiplexed scheme for applying the reservoir computing paradigm on a dynamical system with delayed feedback.
In this case, the measured states for one input-output pair reaction $\vec{s}_n = (s_{1n},s_{2n},\ldots,s_{Mn})$ are recorded at different times $t_m = t_n + m\theta$, with $m = 1,2,\ldots,M$, where $t_n$ is the time at which the $n$-th input $u_n$ is fed into the system. $\theta$ is describing the distance between two recorded states of the system and is called the virtual node separation time.
The time between two inputs $t_{n+1} - t_n$ is called the clock cycle $T$ and describes the period length in which one input $u_n$ is applied to the system.
To get different reactions between two virtual nodes a time-multiplexed masking process is applied. The information fed into the system is preprocessed by multiplying a $T$-periodic mask $g$ on the inputs (see sketch Fig. \ref{fig:sketch}), which is a piecewise constant function consisting of $M$ intervalls, each of length $\theta$.
This corresponds to the input weights in the spatially extended system with the difference that now the input weights are distributed over time. \\
Dambre et al. showed in \cite{DAM12} that the computational capability of a system can be quantified completely via a complete orthonormal set of basis functions on a sequence of inputs $\bold{u}_n = (\ldots,u_{n-2}, u_{n-1}, u_n)$ at time $n$. In this case the index indicates the input $\Delta n$ time steps ago.
The goal is to investigate how the system transforms the inputs $\bold{u}_n$. For this the chosen basis functions $z(\bold{u}_n)$, forming a Hilbert space, are constructed and used to describe every possible transformation on the inputs $\bold{u}_n$. The system's capability to approximate these basis functions is evaluated.
%This means the method analyses the performance of the system to approximate a full basis of transformations on the inputs $\vec{u}$. \\
Consider the following examples: The function $z(\bold{u}_n) = u_{n-5}$ is chosen as a task $o_n$. This is a transformation of the input sequence 5 steps back. The question this task asks is, how well the system can remember the input 5 steps ago.
Another case would be $o_n = z(\bold{u}_n) = u_{n-5}u_{n-2}$, asking how well it can perform the nonlinear transformation of multiplying the input 5 steps into the past with the input 2 steps into the past.
A useful quantity to measure the capability of the system is the capacity defined as
\begin{align}
    \text{C} = 1 - \text{NRMSE}^2.
    \label{eq:capacity}
\end{align}
A value of 1 corresponds to the system being perfectly capable of approximating the transformation task and 0 corresponds to no capability at all. 
A simpler method, giving equal results like Eq. (\ref{eq:capacity}) developed by Dambre et al. in \cite{DAM12} to calculate C is given by
\begin{align}
    \text{C} = \frac{\vec{o} \doubleunderline{s} (\doubleunderline{s}^T\doubleunderline{s})^{-1} \doubleunderline{s}^T \vec{o} }{\norm{\vec{o}}^2},
    \label{eq:mpsi_mem_capacity}
\end{align}
where $^T$ indicates the transpose of a matrix and $^{-1}$ the inverse. We use Eq. (\ref{eq:mpsi_mem_capacity}) to calculate the memory capacity.\\
In this paper we use finite products of normalized Legendre polynomials $P_{d_{\Delta n}}$ as a full basis of the constructed Hilbert space for each input step combination. 
$d$ is the order of the used Legendre polynomial and $\Delta n$ the $\Delta n$-th step into the past passed as value to the Legendre polynomial. Multiplying a set of those Legendre polynomials gives the target task $y_{\{d_{\Delta n}\}}$, which yields (see example below for clarification)
\begin{align}
    y_{\{d_{\Delta n}\}} = \Pi_{\Delta n} P_{d_{\Delta n}}(u_{-\Delta n}).
    \label{eq:LP_construction}
\end{align}
This is directly taken from \cite{DAM12}.
It is important that the inputs to the system are uniformly distributed random numbers $u_n$, which are independent and identically drawn in $[-1,1]$ to match the used normalized Legendre polynomials.
To calculate the memory capacity $MC^d$ for a degree $d$, a summation over all possible past input sets is done
\begin{align}
    MC^d = \sum_{\{\Delta n\}}C_{\{\Delta n\}}^d,
    \label{eq:memory_capacity}
\end{align}
where $\{\Delta n\}$ is the set of past input steps, $C_{\{\Delta n\}}^d$ is the capacity of the system to approximate a specific transformation task $z_{\{\Delta n\}}(\mathbf{u}_n)$ and $d$ is the degree of all Legendre polynomials combined in the task $z_{\{\Delta n\}}(\mathbf{u}_n)$. In the example from above with $z_{\{-5,-2\}}(\mathbf{u}_n) = u_{n-5}u_{n-2}$, it is $d=2$ and $\{\Delta n\}=\{-5,-2\}$.
For $d=1$ we get the well known linear memory capacity.
To compute the total memory capacity, a summation over all degrees $d$ is done. 
\begin{align}
    MC = \sum_{d=1}^D MC^d
    \label{eq:total_mem_cap}
\end{align}%
Dambre et al. showed in \cite{DAM12} that the $MC$ is limited by the readout-dimension $M$, given here by the number of virtual nodes $N_V$. \\
The simulation was written in C$++$ with standard libraries used except for linear algebra calculations, which were calculated via the library "Armadillo".
A Runge-Kutta 4th order method was applied to integrate numerically the delay-differential equation given by Eq. (\ref{eq:stuart_landau}) with an integration step $\Delta t=0.01$ in time units of the system.
First, the system was simulated without any inputs to let transients decay. Afterwards a buffer time was applied with 100000 inputs, that were excluded from the training process. Then, the training and testing process itself was done with 250000 inputs to have sufficient statistics.
The tasks are constructed via Eq. (\ref{eq:LP_construction}) and the corresponding capacities $C_{\{\Delta n\}}^d$ were calcualated via Eq. (\ref{eq:mpsi_mem_capacity}). 
All possible combinations of the Legendre polyomials up to degree $D=10$ and $\Delta n=1000$ input steps into the past were considered. $C_{\{\Delta n\}}^d$ below $0.001$ were excluded because of finite statistics. To calcuate the inverse, the Moore–Penrose pseudoinverse from the C++ linear algebra library "Armadillo" was used. \\% It is important that the random numbers $u$ are independent and identically drawn in $[-1,1]$ to match the normalized Legendre polynomials. \\
We characterize the performance of our nonlinear oscillator by evaluating the total memory capacity $MC$, the contributions $MC^d$ as well as the NRMSE of the NARMA10 task. The latter is a benchmark test and combines memory and nonlinear transformations. It is given by an iterative formula
\begin{align}
    A_{n+1} =  0.3A_n + 0.05A_n \left( \sum_{i=0}^{9}A_{n-i} \right) + 1.5 u_{n-9}u_n + 0.1.
\end{align}
Here, $A_n$ is an iteratively given number and $u_n$ is an independent and identically drawn uniformly distributed random number in $[0,0.5]$.
The reservoir is driven by the random numbers $u_n$ and has to be able to predict the value of $A_{n+1}$, $\vec{o} = \vec{A}$.
\begin{table}%
\begin{center}%
\caption{Parameters used in the simulation if not stated otherwise.}
\begin{adjustbox}{max width=\textwidth}
    \begin{tabular}{ c c c }%
    \hline
    \hline
    \textbf{Parameter} & \textbf{Description} & \textbf{Value}  \\ \hline
    $\lambda$ & pump rate & $-0.02$  \\ \hline
    $\eta$ & input strength & $0.01$  \\ \hline
    $\omega$ & free running frequency & $0.0$  \\ \hline
    $\gamma$ & nonlinearity & $-0.1$  \\ \hline
    $\kappa$ & feedback strength & $0.1$  \\ \hline
    $\theta$ & feedback phase & $0.0$  \\ \hline
    $N_V$ & Number of virtual nodes & $50$  \\ \hline
    $T$ & Input period time & $80$  \\ \hline\hline
    \end{tabular}%
    \end{adjustbox}%
   % \newline
    \label{table}
\end{center}%
\end{table}%
The reservoir we use for our analysis is a Stuart-Landau oscillator, also called Hopf normal form \cite{SCH07}, with delayed feedback. 
This is a generalized model applicable for all systems operated close to a Hopf bifurcation, i.e. close to the onset of intensity oscillations. One example would be a laser operated closely above threshold \cite{ERN10b}. A derivation from the Class B rate equations is shown in the Appendix. 
The equation of motion is given by
\begin{align}
    \dot{Z} = (\lambda + \eta gI + i \omega + \gamma |Z|^2)Z + \kappa e^{i \phi}Z(t - \tau) ,
    \label{eq:stuart_landau}
\end{align}
and was taken from \cite{ROE18a}.
Here, $Z$ is a complex dynamical variable (in the case of a laser $|Z|^2$ resembles the intensity), $\lambda$ is a dimensionless pump rate, $\eta$ the input strength of the information fed into the system via eletrical injection, $g$ is the masking function, $I$ is the input,  $\omega$ is the frequency with which the dynamical variable $Z$ rotates in the complex plane without feedback (in case of a laser, this is the frequency of the emitted laser light), $\gamma$ the nonlinearity in the system, $\kappa$ is the feedback strength, $\phi$ the feedback phase and $\tau$ the delay time. The corresponding parameters used in the simulations are found in Tab. \ref{table} if not stated otherwise.

\section{Results}
\label{Results}
\begin{figure}%
    \centering
    \includegraphics[width=\textwidth]{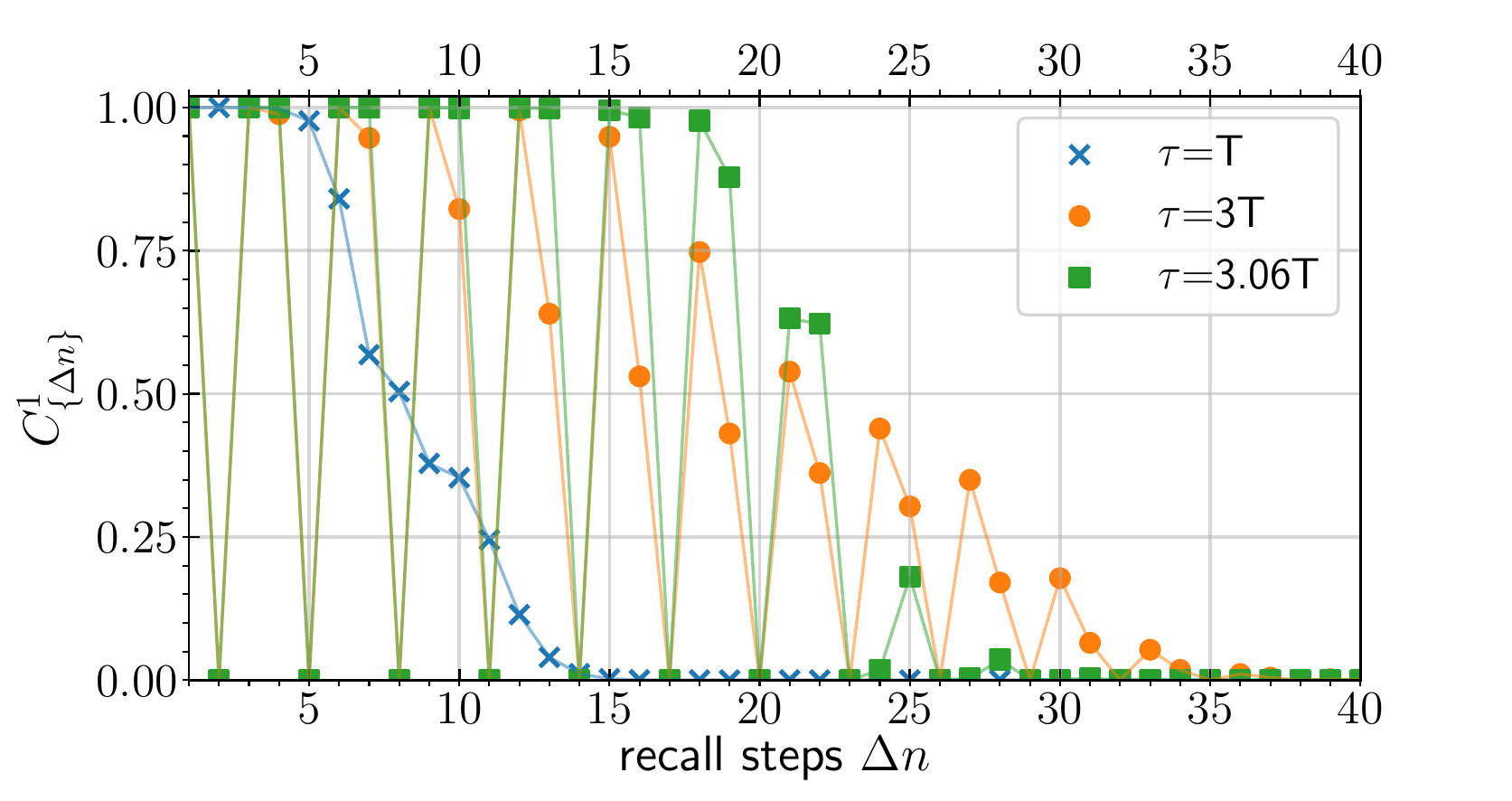}
    \caption{$C_{\{\Delta n\}}^1$ as defined in Eq. \ref{eq:memory_capacity} plotted over the $\Delta n$-th input step to recall for 3 different delay times $\tau$. The input period time $T=80$.}%
    \label{fig:lin_cap}%
\end{figure}

\begin{figure}%
    \centering
    \includegraphics[width=\textwidth]{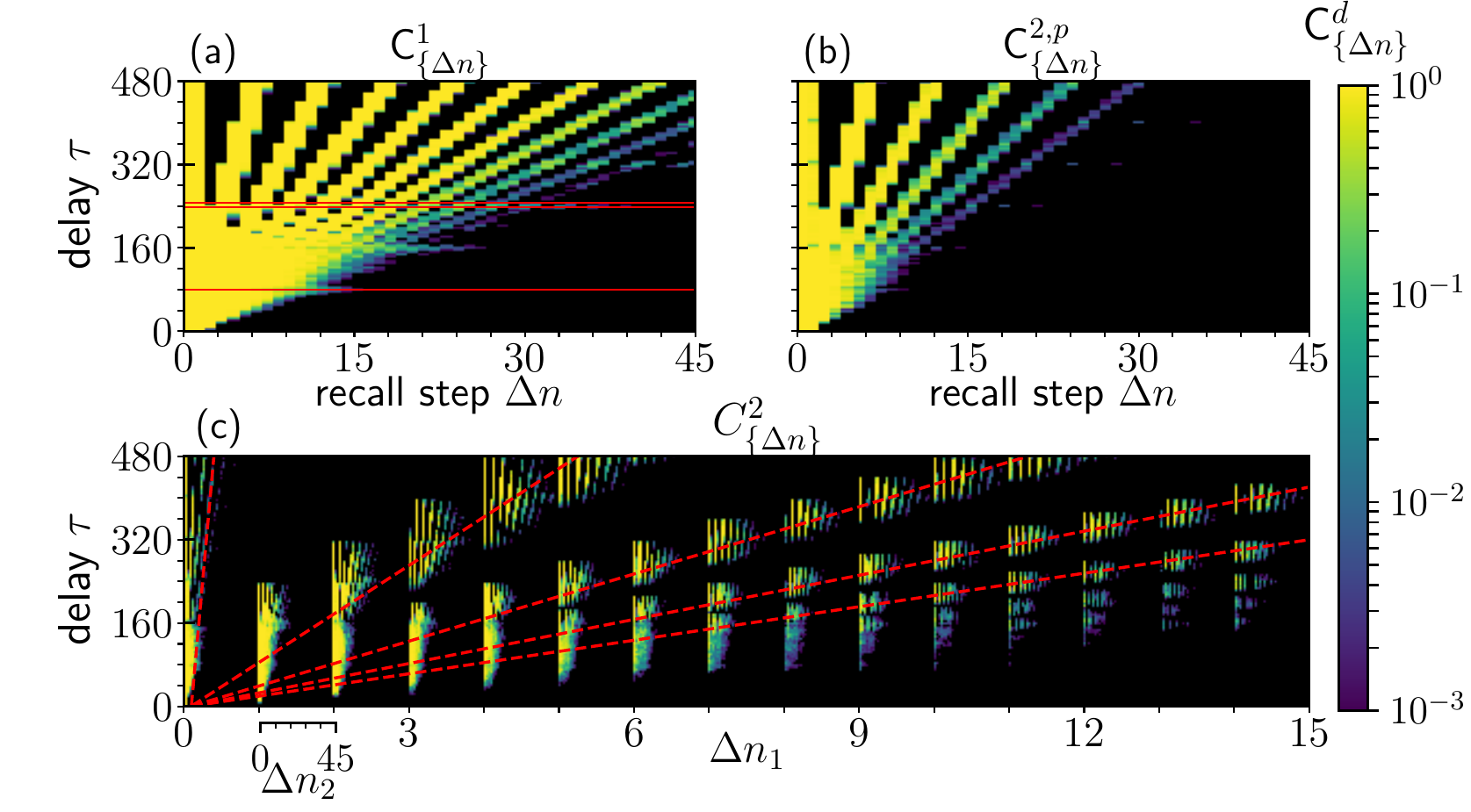}
    \caption{(a) Linear capacity $C^1_{\{\Delta n\}}$ plotted colorcoded over the delay time $\tau$ and the input steps $\Delta n$ to recall. Parameters as given in Tab. \ref{table}. The red horizontal solid lines indicate the scan from Fig. \ref{fig:lin_cap}.  (b) Quadratic pure capacity $C^{2,p}_{\{\Delta n\}}$. (c) Combination of two legendre polynomials of degree 1 indicating the capabiltiy of nonlinear transformations of the form $u_{-\Delta n_1}u_{-\Delta n_2}$. Here $\Delta n_1$ of the first polynomial is plotted while between two $\Delta n_1$-steps $\Delta n_2$ is increased from $0$ to $45$ steps into the past. Yellow indicates good while blue and black indicates bad recallability. The input period time $T=80$.}%
    \label{fig:mem_single_steps}%
\end{figure}

\begin{figure}%
    \centering
    \includegraphics[width=\textwidth]{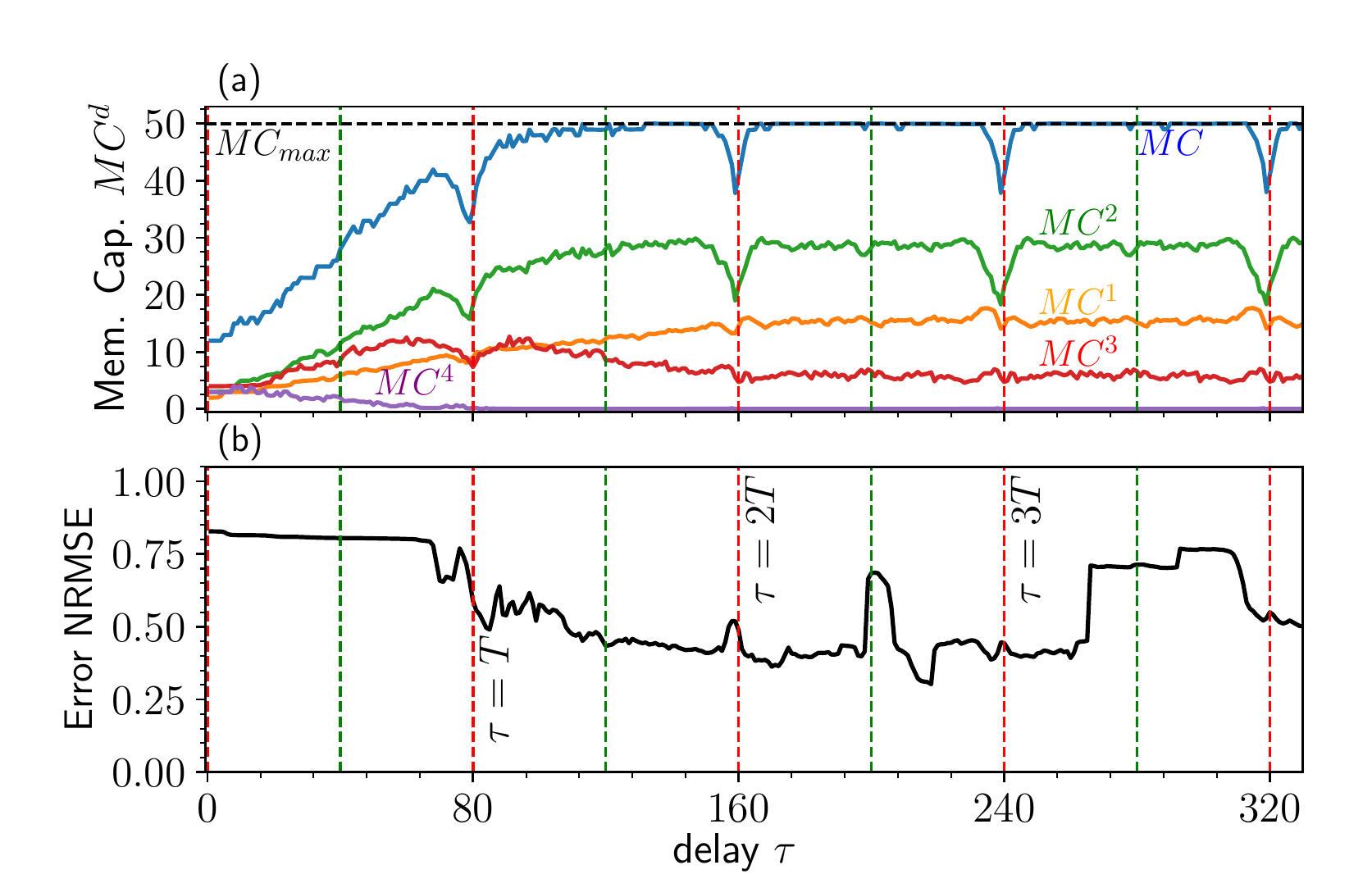}
    \caption{Total memory capacity $MC$ as defined by Eq. \ref{eq:total_mem_cap} (blue) and memory capacities $MC^{1,2,3,4}$ of degree 1 to 4 (orange, green, red, violet)  plotted over the delay time $\tau$ for the same parameters as in Fig. \ref{fig:mem_single_steps}. Resonances between the clock cycle $T$ and the delay time $\tau$ are depicted as vertical red and green dashed lines. One can see the loss in memory capacity at the resonances, especially for degree 2. Higher order transformations with $d>3$ are more effective in the regime where $\tau<1.5\,T$.}%
    \label{fig:total_memory}%
\end{figure}

To get a first impression about how the system can recall inputs from the input sequence \\ $\mathbf{u}_n = (...u_{n-2}, u_{n-1},u_{n})$, we show the linear recall capacities $C_{\{\Delta n\}}^1$ in Fig. \ref{fig:lin_cap}. 
Here, each set of all inputs $\{\Delta n\}$ consists of only one input step $\Delta n$, because $d=1$ for which $z_{\{\Delta n\}}(\mathbf{u}_n)$ consists of the Legendre polynomial $P_1(u_{n-\Delta n}) = u_{n-\Delta n}$.
The capacities $C_{\{\Delta n\}}^1$ are plotted over the step $\Delta n$ to be recalled for 3 different delay times $\tau$ (blue, orange and green in Fig. \ref{fig:lin_cap}) while the input period time $T$ is kept fixed to 80 and the readout dimensions $N_V$ to 50. These timescale parameters were chosen to fit the characteristic timescale of the system, such that the time between two virtual nodes $\theta$ is long enough for the system to react, but short enough such that the speeding process is still as high as possible. For input period times $T=\tau$ (the blue solid line in Fig. \ref{fig:lin_cap}) a high capacity is achieved for a few recall steps after which the recallability drops steadily down to 0 at about the 15-th step ($\Delta n=15$) to recall.
This changes when the input period time reaches values of 3 times the delay time $\tau=3T$ (the orange solid line in Fig. \ref{fig:lin_cap}). Here, the linear recallability $C_{\{\Delta n\}}^1$ oscillates between high and low values as a function of $n$, while its envelope steadily decreases until it reaches 0 at around the 35-th ($\Delta n=35$) step to be recalled. Considering that $\tau=3T$ is a resonance between the input period time $T$ and the delay time $\tau$, one can also take a look at the case for off-resonant setups, which is shown by the green solid line in Fig. \ref{fig:lin_cap} with $\tau \approx3.06T$. 
This parameter choice shows a similar behavior as the $T=3\tau$ one but with higher capacities for short recall steps and a faster decay of the recallability at around the 29-th ($\Delta n=29$) step. \\
To get a more complete picture, we evaluated the linear capacity $C_{\{\Delta n\}}^1$ and quadratic capacities $C_{\{\Delta n\}}^2$ of the system and depicted these as a heatmap over the delay time $\tau$ and the input steps in Fig. \ref{fig:mem_single_steps} for a constant input period time $T$.
The x-axis indicates the $\Delta n$-th step to be recalled while the delay time $\tau$ is varied from bottom to top on the y-axis.
In Fig. \ref{fig:mem_single_steps}(a) the linear capacities $C^1_{\{\Delta n\}}$ are shown, for which the red horizontal solid lines indicate the scan from Fig. \ref{fig:lin_cap}
One can see a continuous capacity $C^1_{\{\Delta n\}}$ for $\tau < 2T$ which forks into rays of certain recallable steps $\Delta n$ that linearly increase with the delay time $\tau$.
%and in the case of the quadratic MC to perform a nonlinear transformation on the inputs.
%For short time delays $\tau$ the system is capable of continously remembering a few inputs into the past which drops fast to values close to 0. Steps of only 3 to 4 inputs with a time delay $\tau$ of less than 40 and up to 20 steps for a time delay at around $\tau=160$ can be recalled.
%After around $\tau=2T$ the continous linear memory recallability forks into rays for certain recallable steps that increase linearly with the delay time $\tau$. 
This implies that specific steps $\Delta n$ can be remembered while others inbetween are forgotten, a crucial limitation to the performance of the system. Generally the number of steps into the past that can be remembered increases with $\tau$(at constant $T$), while on the other hand also the gaps inbetween the recallable steps increase. Thus, the total memory capacity stays constant. This will be discussed later in Fig. \ref{fig:total_memory}.
In Fig. \ref{fig:mem_single_steps}(b) the pure quadratic capacity $C^{2,p}_{\{\Delta n\}}$ is plotted within the same parameter space as in the Fig. \ref{fig:mem_single_steps}(a). 
Pure means that only Legendre polynomials of degree 2, i.e. $P_2(u_{n-\Delta n}) = \frac{1}{2}(3u_{n-\Delta n}^2 - 1)$ were considered, rather than also considering combinations of two Legendre polynomials of degree 1, i.e. $P_1(u_{n-\Delta n_1})P_1(u_{n-\Delta n_2}) = u_{n-\Delta n_1}u_{n-\Delta n_2}$.
In the graph one can see the same behaviour as for the linear capacities $C^1_{\{\Delta n\}}$ (Fig. \ref{fig:mem_single_steps}(a)), but with less rays and thus less steps that can be remembered from the past.
This indicates that the dynamical system is not as effective in recalling inputs and additionally transforming them nonlinearily as it is in just recalling them linearily.
For the full quadratic nonlinear transformation capacity, all combinations of two Legendre polynomials of degree 1 for different input steps into the past have to be considered, i.e. \\$P_1(u_{n-\Delta n_1})P_1(u_{n-\Delta n_2}) = u_{n-\Delta n_1}u_{n-\Delta n_2}$. \\
This is shown in Fig. \ref{fig:mem_single_steps}(c).
Again, the capacities $C^2_{\{\Delta n\}}$ are depicted as a heatmap and the delay time $\tau$ is varied along the y-axis. 
This time the x-axis shows the steps of the first Legendre polynomial $\Delta n_1$, while inbetween two ticks of the x-axis, the second Legendre polynomial's step $\Delta n_2$ is scanned from 0 up to 45 steps into the past.
For the steps of the second Legendre polynomial $\Delta n_2$ the capacity exhibits the same behaviour as already discussed for Fig. \ref{fig:mem_single_steps}(a) and (b). This does also apply to the first Legendre polynomial which induces interference patterns in the capacity space of the two combined legendre polynomials. The red dashed lines highlight the ray behaviour of the first Legendre polynomial. 
%If one concatenates every step of the first legendre polynomial, while keeping the step of the second constant (in our case 0), one gets the same behaviour as in the upper plots.
We therefore learn that the performance of a reservoir computer described by a Hopf normal form with delay drastically depends on the task.
There are certain nonlinear transformation combinations $u_{-\Delta n_1}u_{-\Delta n_2}$ of the inputs $u_{-\Delta n_1}$ and $u_{-\Delta n_2}$ which cannot be approximated due to the missing memory at specific steps. To overcome these limitations it would be recommended to use multiple systems with different parameters to compensate for each other. \\
To fully characterize the computational capabilities of our reservoir computer, a full analysis of the degree $d$ memory capacities $MC^d$ and the total memory capacity $MC$ as defined in Eq. (\ref{eq:total_mem_cap}) is done.
%$MC^d$ is computed by summing over all capacities for $d$, e.g. $MC^1$ is computed by summation along the x-axis in Fig.\ref{fig:mem_single_steps}(a) and $MC^2$ by summation along the x-axis for both Fig. \ref{fig:mem_single_steps}(b) and Fig. \ref{fig:mem_single_steps}(c) combined, while not double counting ($u_{-n_i}u_{-n_j}$ = $u_{-n_j}u_{-n_i}$). 
%Total memory capacity $MC$ is the summation of all memory capacities $MC^d$. 
\begin{figure*}[!htb]%
    \centering
    \includegraphics[width=0.75\textwidth]{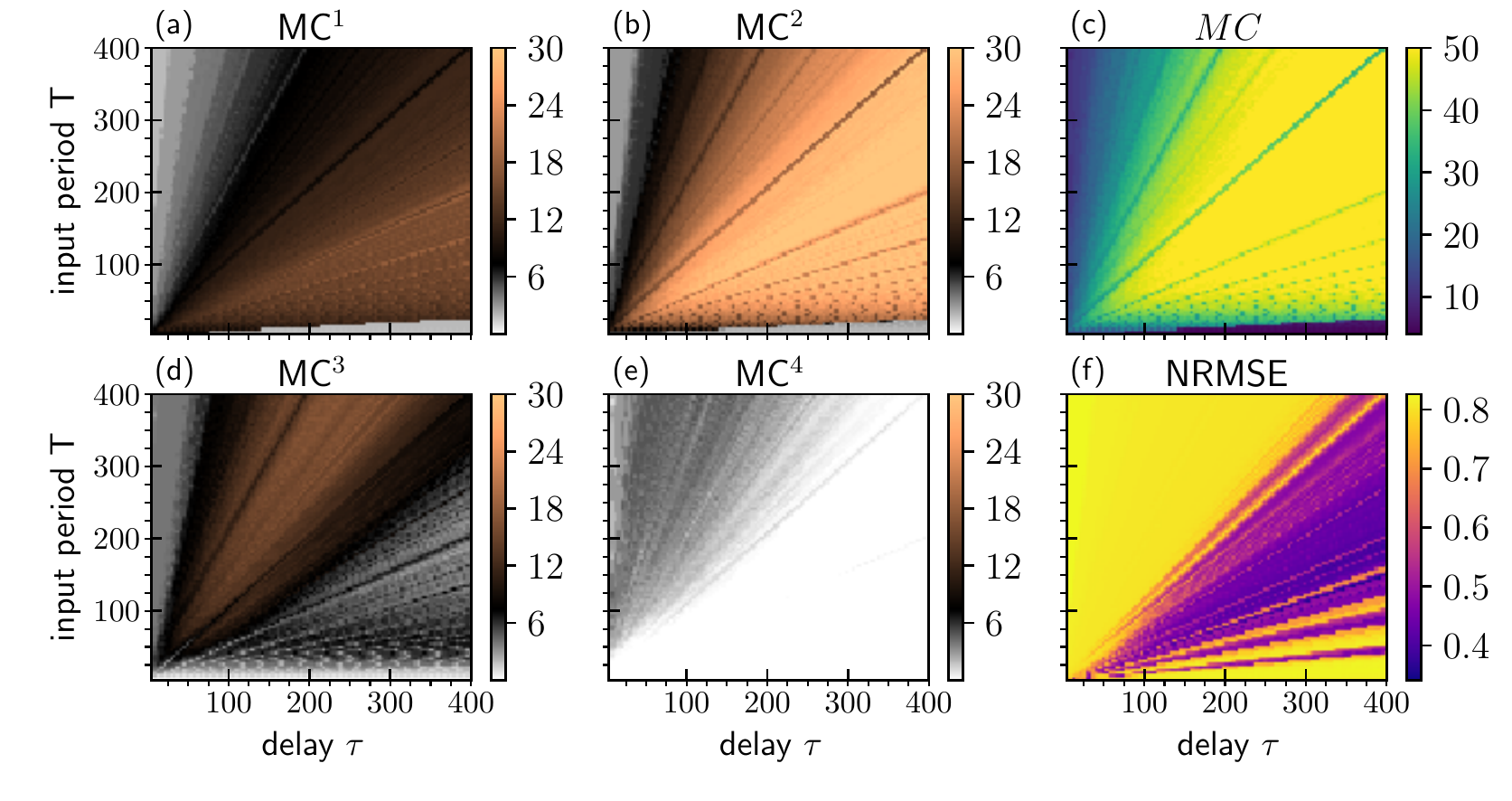}
    \caption{(a) Linear memory capacity MC$^1$ plotted colorcoded over the delay time $\tau$ and the input period time $T$. (b) Degree 2, MC$^2$. (c) Total memory capacity, MC. (d) Degree 3, MC$^3$. (e) Degree 4, $MC^4$. (f) NARMA10 prediction error NRMSE.}%
    \label{fig:2D-scan}%
\end{figure*}%
The results are depicted in Fig. \ref{fig:total_memory}(a) as a function of the delay time $\tau$. All other parameters are fixed as in Fig. \ref{fig:mem_single_steps}.
The orange solid line in Fig. \ref{fig:total_memory} referes to the linear, the green, red and violet lines to the quadratic, cubic and quartic memory capacity $MC^{1,2,3,4}$, respectively. The blue solid line shows the total memory capacity $MC$ summed up over all degrees up to 10. Dambre et al. showed in \cite{DAM12} that the $MC$ is limited by the number of read-out dimensions and equals it when all read-out dimensions are linearly independent. In our case the read-out dimension is given by the number of virtual nodes $N_V=50$.
Nevertheless, the total memory capacity $MC$ starts at around 15 for a very short time delay with respect to the input period time $T$. This low value arises from the fact that a short delay induces a high correlation between the responses of the dynamical system which induces highly linearly dependent virtual nodes. 
This is an important general result that has to be kept in mind for all delay based reservoir computing systems: With $\tau<1.5\,T$ the capability of the reservoir computer is partially waisted.
Increasing the delay time $\tau$ also increases the total memory capacity $MC$ reaching the upper bound of 50 at around $1.5$ times the input period time $T$. \\
For $\tau>1.5\,T$ an interesting behaviour emerges. Depicted by the vertical red dashed lines are multiples of the input period time $T$ at which the total memory capacity $MC$ drops again significantly to around 40. A drop in the linear memory capacity was discussed in the paper by Stelzer et al. \cite{STE20} and explained by the fact that resonances between the delay time $\tau$ and the input period time $T$ concludes in a sparse connection between the virtual nodes.
Our results now show that this effects the total memory capacity $MC$, by mainly reducing the quadratic memory capacity $MC^2$. At the resonances the quadratic nonlinear transformation capability of the system is reduced. 
To conclude, delay based reservoir computing systems should be kept off the resonances between  $T$ and $\tau$ to maximize the computational capability.
%that has a highly quadratic nonlinear transformation part 
A suprising result is that for the chosen Hopf nonlinearity the linear memory capacity $MC^1$ is only slightly influenced by the resonances. 
A result from Dambre et al. in \cite{DAM12} and analysed by Inubushi et al. in \cite{INU17} showed that a trade-off between the linear recallability and the nonlinear transformation capability exists. 
This is clearly only the case if the theoretical limit of the total memory capacity $MC$ is reached and kept constant, thus every change in the linear memory capacity $MC^1$ has to induce a change in the nonlinear memory capacities $MC^d$, $d>1$. In the case of resonances, a decrease in the total memory capacity $MC$ happens and thus this loss can be distributed in any possible way over the different memory capacities $MC^d$. In our case, we see that the influence on the quadratic memory capacity $MC^2$ is highest. \\
The system is capable of a small amount of cubic transformations, depicted by the solid red line in Fig. \ref{fig:total_memory}(a), which also decreases at the resonances in a similar way as the quadratic contribution does.
Higher order memory capacities $MC^d$, with $d>3$, have only small contributions for short delay times $\tau$, dropping to 0 for increased time delay $\tau$. 
A possible explanation is the fact that short delays induce an interaction of the last input directly with itself for $k=\frac{T}{\tau}$ times, depending on the ratio between $\tau$ and $T$.
As a result, short delay times $\tau$ enable highly nonlinear tasks in expense of a lower total memory capacity $MC$. \\
For more insights into the computing capabilities of our nonlinaer oscillator we now also discuss the NARMA10 time series prediction task, shown in Fig. \ref{fig:total_memory}(b).
Comparing the memory capacities $MC^d$ to the NARMA10 computation error NRMSE in \ref{fig:total_memory}(b), a small increase in the NARMA10 NRMSE can be seen at the resonances with $n\tau = mT$, where $n \in [0,1,2...]$ and $m \in [0,1,2...]$.
For a systematic characterization a scan of the input period time $T$ and the delay time $\tau$ was done and the total memory capacity $MC$ (Fig. \ref{fig:2D-scan}(c)), the memory capacities of degrees 1-4 $MC^{1,2,3,4}$ (Fig. \ref{fig:2D-scan}(a,b,d,e)) and the NARMA10 NRMSE (Fig. \ref{fig:2D-scan}(f)) were plotted colorcoded over the two timescales. This is an extension of the results of Röhm et al. in \cite{ROE20}, where only the linear memory capacity and the NARMA10 computation error were analysed.
For short time delays $\tau$ and period input times $T$ the memory capacities of degree 1-3 $MC^{1,2,3}$ and the total memory capacity $MC$ are significantly below the theoretical limit of 50 as already seen in the results from Fig. \ref{fig:total_memory}, while the NARMA10 NRMSE also has high errors of around $0.8$. This comes from the fact that short input period times $T$ also mean short virtual node distances $\theta$, which induces a high linear correlation between the read-out dimensions.
Degree 4 on the other hand only has values as long as $T>\tau$, a result coming from the fact that the input $u_n$ has to interact with itself to get a transformation of degree 4. A possible explanation comes from the fact that the dynamical system itself is not capable of transformations higher than degree 3, since the highest order in Eq. (\ref{eq:stuart_landau}) is 3.
If the delay time $\tau$ and the input period time $T$ are long enough the total memory capacity $MC$ reaches 50 with exceptions of resonances between $\tau$ and $T$. 
These resonances are also seen in the NARMA10 NRMSE for which higher errors occur.
Looking at the memory capacity of degree 1 and 2 $MC^{1,2}$ and comparing it with the NARMA10 NRMSE one can see a tendency in which the NARMA10 NRMSE is lowest where both have the highest capacities, raising from the fact that the NARMA10 task is highly dependent on linear memory and quadratic nonlinear transformations.
This can also be seen in the area below the $\tau=T$-resonance. 
To conclude, one can use the parameter dependencies of the memory capacities $MC^d$ to make predictions of the reservoirs capability to approximate certain tasks.\\
%In general we have analysed the memory capacities of an oscillatory system with delayed feedback operated close to a Hopf bifurcation. Particularly, a recallability behaviour was found for the system in which only certain input steps could be rememberd, while jumps between the recallable steps are found. This behaviour was also found for higher degrees of MCs.
%An analysis of the summed MCs for every degree showed a significant influence of resonances between $\tau$ and $T$, especially for the quadratic MCs.

\section{Conclusions}
\label{Conclusions}
We analysed the memory capacities and nonlinear transformation capabilities of a reservoir computer consisting of an oscillatory system with delayed feedback operated close to a Hopf bifurcation, i.e. a paradigmatic model also applicable for lasers close to threshold. We systematically varied the timescales and found regions of high and low reservoir computing performing abilities.
Resonances between the information input period time $T$ and the delay time $\tau$ should be avoided to fully utilize the natural computational capability of the nonlinear oscillator. A ratio of $\tau=1.6T$ was found to be the optimal for the computed memory capacities, resulting in a good NARMA10 task approximation. Furthermore, it was shown, that the recallability for high delay times $\tau \gg T$ is restricted to specific past inputs, which rules out certain tasks. By computing the memory capacities of a Hopf normal form, one can make general assumptions about the reservoir computing capabilities of any system operated close to a Hopf bifurcation. This significantly helps in understanding and predicting the task dependence of reservoir computers.

%This restriction shows up as a ray structure in the parameter space of recalled steps $n$ and delay time $\tau$.

\begin{acknowledgements}
The authors would like to thank André Röhm, Joni Dambre and David Hering for fruitfull discussion.
%The authors also acknowledge the financial support by the "Deutsche Forschungsgemeinschaft" (DFG) in the framework of SFB 910.
\end{acknowledgements}

\par\addvspace{17pt} \noindent \small\rmfamily{\textbf{Funding Information} This study was funded by the "Deutsche Forschungsgemeinschaft" (DFG) in the framework of SFB910.}

\par\addvspace{17pt} \noindent \small\rmfamily{\textbf{Conflict} The Authors declare that they have no conflict of interest.}

\par\addvspace{17pt} \noindent \small\rmfamily{\textbf{Ethical approval} This article does not contain any studies with human participants or animals performed by any of the authors.}

%\bibliographystyle{spbasic.bst}      % basic style, author-year citations
\bibliographystyle{spphys}       % APS-like style for physics
\bibliography{ref.bib}

%\begin{thebibliography}{}
% etc
%\end{thebibliography}

\section{Appendix}

Derivation of the Stuart-Landau Equation with delay from the Class-B laser rate equations
\begin{align}
    \dot{E} &= (1 + i \alpha)EN \label{eq:LK} \\
    \dot{N} &= \frac{1}{T}(P + \eta g I - N - (1+2N)|E|^2),
\end{align}
where $E$ is the non-dimensionalized complex eletrical field and $N$ the non-dimensionalized carrier inversion, P the pump relativ to the threshold for $P_{thresh} = 0$ and $\alpha$ the Henry factor.
The reservoir computing signal is fed into the system via electrical injection $\eta g I$.
If fast carriers are considered, an adiabatic elimination of the charge carriers yields
\begin{align}
0 &= \frac{1}{T}(P + \eta g I - N - (1+2N)|E|^2) \\
N &= \frac{P + \eta g I - |E|^2}{1+2|E|^2}
\end{align}
,which after substituting into Eq. (\ref{eq:LK}) gives
\begin{align}
    \dot{E} = (1 + i \alpha)E\frac{\tilde{P} - |E|^2}{1+2|E|^2},
\end{align}
where we introduced the quantity $\tilde{P} = P + \eta g I$ for convenience purposes.
This equation yields the full Class A rate equation for the non-dimensionalized complex electric field.
Simulations with the full Class A rate equation close to the threshold show similar results to the reduced case. \\
Because we consider laser that are operated close to the threshold level, a taylor expansion of the denominator for $|E|^2\approx 0$ is done, 
\begin{align}
    \dot{E} = (1 + i \alpha)E (\tilde{P} - |E|^2- 2\tilde{P}|E|^2) ,
\end{align}
where we set $|E|^4 \approx 0$ as a neglectable term.
As we consider a laser operated close to the threshold, it follows that the pump $\tilde{P}$ and the intensity $|E|^2$ are of Order $O(\epsilon)$, where $\epsilon$ is a small factor. This holds true only if the input signal $\eta g I$ is a small electrical injection. After applying this the equation is given by
\begin{align}
    \dot{E} = (1 + i \alpha)E (\tilde{P} - |E|^2)
\end{align}
We can substitute $\tilde{P} = P + \eta g I$ back into the equation, change the rotating frame of the laser by setting $E = Z e^{-i(\omega - \alpha \tilde{P}) t}$ and introduce a complex factor $\gamma=-(1 + i \alpha)$ that scales the nonlinearity 
\begin{align}
    \dot{Z} = Z (P + \eta g I + i\omega + \gamma|Z|^2),
\end{align}
By addding feedback $\kappa e^{\phi}Z(t-\tau)$ to the system one arrives at Eq. (\ref{eq:stuart_landau}).
\begin{align}
    \dot{Z} = Z (P + \eta g I + i\omega + \gamma|Z|^2) + \kappa e^{\phi}Z(t-\tau),
\end{align}
\end{document}